\documentstyle[11pt]{article}

\def\ap{a^{\dagger}}
\def\alg{${\cal A}^{(\lambda)}_{\alpha_0 \alpha_1 \ldots \alpha_{\lambda-2}}$}
\def\algtwo{${\cal A}^{(2)}_{\alpha_0}$}
\def\gdoa{${\cal A}^{(\lambda)}(G(N))$}
\def\gdoatwo{${\cal A}^{(2)}(G(N))$}
\def\case#1#2{{\textstyle{#1\over #2}}}
\def\Qp{Q^{\dagger}}
\def\Ap{A^{\dagger}}

\title{
C$_\lambda$-extended oscillator algebras: Theory and applications to
(variants of) supersymmetric quantum mechanics}

\author{C.~Quesne\thanks{\rm Directeur de recherches FNRS; E-mail:
cquesne@ulb.ac.be}\ , N.~Vansteenkiste\thanks{\rm E-mail:
nvsteen@ulb.ac.be}\\
{\small \em Physique Nucl\'eaire Th\'eorique et Physique Math\'ematique,
Universit\'e Libre de Bruxelles,}\\
{\small \em Campus de la Plaine CP229, Boulevard~du Triomphe, B-1050 Brussels,
Belgium}}

\date{}

\begin{document}

\maketitle
\begin{abstract}

C$_{\lambda}$-extended oscillator algebras, where C$_{\lambda}$ is the cyclic
group of order~$\lambda$, are introduced and realized as generalized deformed
oscillator algebras. For $\lambda=2$, they reduce to the well-known
Calogero-Vasiliev algebra. For higher $\lambda$ values, they are shown to
provide
in their bosonic Fock space representation some interesting applications to
supersymmetric quantum mechanics and some variants thereof: an algebraic
realization of supersymmetric quantum mechanics for cyclic shape invariant
potentials of period~$\lambda$, a bosonization of parasupersymmetric quantum
mechanics of order $p = \lambda-1$, and, for $\lambda=3$, a bosonization of
pseudosupersymmetric quantum mechanics and orthosupersymmetric quantum
mechanics of order two.

\end{abstract}
%
%
\section{Introduction}

Deformations and extensions of the oscillator algebra have found a lot of
applications to physical problems, such as the description of systems with
non-standard statistics, the construction of integrable lattice models, the
investigation of nonlinearities in quantum optics, as well as the algebraic
treatment of quantum exactly solvable models and of $n$-particle integrable
systems.\par
%
%
The generalized deformed oscillator algebras (GDOAs) (see e.g.\ Ref.\
\cite{cq95}
and references quoted therein) arose from successive generalizations of the
Arik-Coon~\cite{arik} and Biedenharn-Macfarlane~\cite{biedenharn}
$q$-oscillators. Such algebras, denoted by ${\cal A}_q(G(N))$, are
generated by the
unit, creation, annihilation, and number operators $I$, $\ap$, $a$, $N$, satisfying
the Hermiticity conditions $\left(\ap\right)^{\dagger} = a$, $N^{\dagger} =
N$, and
the commutation relations
\begin{equation}
  \left[N, \ap\right] = \ap, \qquad [N, a] = - a, \qquad  \left[a,
\ap\right]_q \equiv
  a \ap - q \ap a = G(N),
\end{equation}
where $q$ is some real number and $G(N)$ is some Hermitian, analytic
function.\par
%
%
On the other hand, $\cal G$-extended oscillator algebras, where $\cal G$ is some
finite group, appeared in connection  with $n$-particle integrable models.
For the
Calogero model~\cite{calogero}, for instance, $\cal G$ is the symmetric group
$S_n$~\cite{poly}.\par
%
%
{}For two particles, the $S_2$-extended oscillator algebra ${\cal
A}^{(2)}_{\kappa}$, where $S_2 = \{\, I, K \mid K^2 = I \,\}$, is generated
by the
operators $I$, $\ap$, $a$, $N$, $K$, subject to the Hermiticity conditions
$\left(\ap\right)^{\dagger} = a$, $N^{\dagger} = N$, $K^{\dagger} = K^{-1}$, and
the relations
\begin{eqnarray}
  \left[N, a^{\dagger}\right] & = & a^{\dagger}, \qquad [N, K] = 0, \qquad
K^2 = I,
          \nonumber \\
  \left[a, a^{\dagger}\right] & = & I + \kappa K \qquad (\kappa \in {\mbox R}),
  \qquad a^{\dagger} K = - K a^{\dagger},
\end{eqnarray}
together with their Hermitian conjugates.\par
%
%
When the $S_2$ generator $K$ is realized in terms of the Klein operator
$(-1)^N$,
${\cal A}^{(2)}_{\kappa}$ becomes a GDOA characterized by $q=1$ and $G(N) = I +
\kappa (-1)^N$, and known as the Calogero-Vasiliev~\cite{vasiliev} or
modified~\cite{brze} oscillator algebra.\par
%
%
The operator $K$ may be alternatively considered as the generator of the cyclic
group $C_2$ of order two, since the latter is isomorphic to $S_2$. By replacing
$C_2$ by the cyclic group of order $\lambda$, $C_{\lambda} = \{\, I, T,
T^2, \ldots,
T^{\lambda-1} \mid T^{\lambda} = I \,\}$, one then gets a new class of $\cal
G$-extended oscillator algebras~\cite{cq98a}, generalizing that describing the
two-particle Calogero model. In the present communication, we will define the
$C_{\lambda}$-extended oscillator algebras, study some of their properties, and
show that they have some interesting applications to supersymmetric quantum
mechanics (SSQM)~\cite{witten} and some of its variants.\par
%
%
\section{\boldmath Definition and properties of $C_{\lambda}$-extended
oscillator algebras}
\setcounter{equation}{0}

Let us consider the algebras generated by the operators $I$, $\ap$, $a$,
$N$, $T$,
satisfying the Hermiticity conditions $\left(\ap\right)^{\dagger} = a$,
$N^{\dagger} = N$, $T^{\dagger} = T^{-1}$, and the relations
\begin{eqnarray}
  \left[N, \ap\right] & = & \ap, \qquad [N, T] = 0, \qquad T^{\lambda} = I,
\nonumber
            \\
  \left[a, \ap\right] & = & I + \sum_{\mu=1}^{\lambda-1} \kappa_{\mu} T^{\mu},
            \qquad \ap T = e^{-{\rm i}2\pi/\lambda}\, T \ap,
\label{eq:alg-def1}
\end{eqnarray}
together with their Hermitian conjugates~\cite{cq98a}. Here $T$ is the
generator of
(a unitary representation of) the cyclic group $C_{\lambda}$ (where $\lambda \in
\{2, 3, 4, \ldots\}$), and $\kappa_{\mu}$, $\mu = 1$, 2,
$\ldots$,~$\lambda-1$, are
some complex parameters restricted by the conditions $\kappa_{\mu}^* =
\kappa_{\lambda - \mu}$ (so that there remain altogether $\lambda-1$
independent real parameters).\par
%
%
$C_{\lambda}$ has $\lambda$ inequivalent, one-dimensional matrix unitary
irreducible representations (unirreps) $\Gamma^{\mu}$, $\mu = 0$, 1,
$\ldots$,~$\lambda-1$, which are such that $\Gamma^{\mu}\left(T^{\nu}\right) =
\exp({\rm i}2\pi \mu \nu/\lambda)$ for any $\nu = 0$, 1, $\ldots$,~$\lambda-1$.
The projection operator on the carrier space of~$\Gamma^{\mu}$ may be written as
\begin{equation}
  P_{\mu} = \frac{1}{\lambda} \sum_{\nu=0}^{\lambda-1}
  e^{-{\rm i}2\pi \mu\nu/\lambda}\, T^{\nu},
\end{equation}
and conversely $T^{\nu}$, $\nu=0$, 1, $\ldots$,~$\lambda-1$, may be expressed in
terms of the $P_{\mu}$'s as
\begin{equation}
  T^{\nu} = \sum_{\mu=0}^{\lambda-1}  e^{{\rm i}2\pi \mu\nu/\lambda} P_{\mu}.
\end{equation}
\par
%
%
The algebra defining relations~(\ref{eq:alg-def1}) may therefore be rewritten in
terms of $I$, $\ap$, $a$, $N$, and~$P_{\mu}^{\vphantom{\dagger}} =
P_{\mu}^{\dagger}$, $\mu=0$, 1, $\ldots$,~$\lambda-1$, as
\begin{eqnarray}
  \left[N, \ap\right] & = & \ap, \qquad \left[N, P_{\mu}\right] = 0, \qquad
            \sum_{\mu=0}^{\lambda-1} P_{\mu} = I, \nonumber \\
  \left[a, \ap\right] & = & I + \sum_{\mu=0}^{\lambda-1} \alpha_{\mu} P_{\mu},
            \qquad \ap P_{\mu} = P_{\mu+1}\, \ap, \qquad P_{\mu} P_{\nu} =
            \delta_{\mu,\nu} P_{\mu},  \label{eq:alg-def2}
\end{eqnarray}
where we use the convention $P_{\mu'} = P_{\mu}$ if $\mu' - \mu = 0\, {\rm
mod}\,
\lambda$ (and similarly for other operators or parameters indexed by $\mu$,
$\mu'$). Equation~(\ref{eq:alg-def2}) depends upon $\lambda$ real parameters
$\alpha_{\mu} = \sum_{\nu=1}^{\lambda-1} \exp({\rm i}2\pi \mu\nu/\lambda)
\kappa_{\nu}$, $\mu=0$, 1, $\ldots$,~$\lambda-1$, restricted by the condition
$\sum_{\mu=0}^{\lambda-1} \alpha_{\mu} = 0$. Hence, we may eliminate one of
them, for instance $\alpha_{\lambda-1}$, and denote $C_{\lambda}$-extended
oscillator algebras by \alg.\par
%
%
The cyclic group generator $T$ and the projection operators $P_{\mu}$ can be
realized in terms of $N$ as
\begin{equation}
  T = e^{{\rm i}2\pi N/\lambda}, \qquad P_{\mu} = \frac{1}{\lambda}
  \sum_{\nu=0}^{\lambda-1} e^{{\rm i}2\pi \nu (N-\mu)/\lambda}, \qquad \mu
= 0, 1,
  \ldots, \lambda-1,    \label{eq:N-realize}
\end{equation}
respectively. With such a choice, \alg\ becomes a GDOA, \gdoa, characterized by
$q=1$ and $G(N) = I + \sum_{\mu=0}^{\lambda-1} \alpha_{\mu} P_{\mu}$, where
$P_{\mu}$ is given in Eq.~(\ref{eq:N-realize}).\par
%
%
{}For any GDOA ${\cal A}_q(G(N))$, one may define a so-called structure
function~$F(N)$, which is the solution of the difference equation $F(N+1) -
q F(N) =
G(N)$, such that $F(0) = 0$~\cite{cq95}. For \gdoa, we find
\begin{equation}
  F(N) = N + \sum_{\mu=0}^{\lambda-1} \beta_{\mu} P_{\mu}, \qquad \beta_0
  \equiv 0, \qquad \beta_{\mu} \equiv \sum_{\nu=0}^{\mu-1} \alpha_{\nu} \quad
  (\mu =1, 2, \ldots, \lambda-1).
\end{equation}
\par
%
%
At this point, it is worth noting that for $\lambda=2$, we obtain $T=K$,
$P_0 = (I +
K)/2$, $P_1 = (I - K)/2$, and $\kappa_1 = \kappa_1^* = \alpha_0 = - \alpha_1 =
\kappa$, so that \algtwo\ coincides with the $S_2$-extended oscillator algebra
${\cal A}^{(2)}_{\kappa}$ and \gdoatwo\ with the Calogero-Vasiliev algebra.\par
%
%
In Ref.~\cite{cq99b}, we showed that \gdoa\ (and more generally \alg) has
only two
different types of unirreps: infinite-dimensional bounded from below
unirreps and
finite-dimensional ones. Among the former, there is the so-called bosonic Fock
space representation, wherein $\ap a = F(N)$ and $a \ap = F(N+1)$. Its
carrier space
$\cal F$ is spanned by the eigenvectors~$|n\rangle$ of the number operator~$N$,
corresponding to the eigenvalues $n=0$, 1, 2,~$\ldots$, where $|0\rangle$ is a
vacuum state, i.e., $a |0\rangle = N|0\rangle = 0$ and $P_{\mu} |0\rangle =
\delta_{\mu,0} |0\rangle$. The eigenvectors can be written as
\begin{equation}
  |n\rangle = {\cal N}_n^{-1/2} \left(\ap\right)^n |0\rangle, \qquad n = 0,
1, 2,
  \ldots,  \label{eq:vectors}
\end{equation}
where ${\cal N}_n = \prod_{i=1}^n F(i)$. The creation and annihilation operators act
upon~$|n\rangle$ in the usual way, i.e.,
\begin{equation}
  \ap |n\rangle = \sqrt{F(n+1)}\, |n+1\rangle, \qquad a |n\rangle =
\sqrt{F(n)}\,
  |n-1\rangle,
\end{equation}
while $P_{\mu}$ projects on the $\mu$th component ${\cal F}_{\mu} \equiv
\{\, |k\lambda + \mu\rangle \mid k = 0, 1, 2, \ldots\,\}$ of the
${\rm Z}_{\lambda}$-graded Fock space ${\cal F} = \sum_{\mu=0}^{\lambda-1}
\oplus {\cal F}_{\mu}$. It is obvious that such a bosonic Fock space
representation
exists if and only if $F(\mu) > 0$ for $\mu=1$, 2, $\ldots$,~$\lambda-1$. This
gives the following restrictions on the algebra parameters~$\alpha_{\mu}$,
\begin{equation}
  \sum_{\nu=0}^{\mu-1} \alpha_{\nu} > - \mu, \qquad \mu = 1, 2, \ldots,
\lambda-1.
  \label{eq:cond-Fock}
\end{equation}
\par
%
%
In the bosonic Fock space representation, we may consider the bosonic oscillator
Hamiltonian, defined as usual by
\begin{equation}
  H_0 \equiv \case{1}{2} \left\{a, \ap\right\}.  \label{eq:H_0}
\end{equation}
It can be rewritten as
\begin{equation}
  H_0 = \ap a + \frac{1}{2} \left(I + \sum_{\mu=0}^{\lambda-1} \alpha_{\mu}
  P_{\mu}\right) = N + \frac{1}{2} I + \sum_{\mu=0}^{\lambda-1} \gamma_{\mu}
  P_{\mu},
\end{equation}
where $\gamma_0 \equiv \frac{1}{2} \alpha_0$ and $\gamma_{\mu} \equiv
\sum_{\nu=0}^{\mu-1} \alpha_{\nu} + \frac{1}{2} \alpha_{\mu}$ for $\mu = 1$, 2,
\ldots,~$\lambda-1$.\par
%
%
The eigenvectors of $H_0$ are the states~$|n\rangle = |k \lambda + \mu\rangle$,
defined in Eq.~(\ref{eq:vectors}), and their eigenvalues are given by
\begin{equation}
  E_{k\lambda+\mu} = k\lambda + \mu + \gamma_{\mu} + \case{1}{2}, \qquad k = 0,
 1, 2, \ldots, \qquad \mu = 0, 1, \ldots, \lambda-1.
\end{equation}
In each ${\cal F}_{\mu}$ subspace of the ${\rm Z}_{\lambda}$-graded Fock
space~$\cal F$, the spectrum of~$H_0$ is therefore harmonic, but the $\lambda$
infinite sets of equally spaced energy levels, corresponding to $\mu=0$, 1,
$\ldots$,~$\lambda-1$, may be shifted with respect to each other by some
amounts depending upon the algebra parameters $\alpha_0$, $\alpha_1$,
$\ldots$,~$\alpha_{\lambda-2}$, through their linear combinations
$\gamma_{\mu}$, $\mu=0$, 1, $\ldots$,~$\lambda-1$.\par
%
%
{}For the Calogero-Vasiliev oscillator, i.e., for $\lambda=2$, the relation
$\gamma_0 = \gamma_1 = \kappa/2$ implies that the spectrum is very simple and
coincides with that of a shifted harmonic oscillator. For $\lambda\ge 3$,
however,
it has a much richer structure. According to the parameter values, it may be
nondegenerate, or may exhibit some ($\nu+1$)-fold degeneracies above some energy
eigenvalue, where $\nu$ may take any value in the set $\{1, 2, \ldots,
\lambda-1\}$.
In Ref.~\cite{cq99a}, we obtained for $\lambda=3$ the complete classification of
nondegenerate, twofold and threefold degenerate spectra in terms of $\alpha_0$
and $\alpha_1$.\par
%
%
In the remaining part of this communication, we will show that the bosonic Fock
space representation of \gdoa\ and the corresponding bosonic oscillator
Hamiltonian $H_0$ have some useful applications to SSQM and some of its
variants.\par
%
%
\section{Application to supersymmetric quantum mechanics with cyclic shape
invariant potentials}
\setcounter{equation}{0}

In SSQM with two supercharges, the supersymmetric Hamiltonian $\cal H$ and the
supercharges $\Qp$, $Q = \left(\Qp\right)^{\dagger}$, satisfy the sqm(2)
superalgebra, defined by the relations
\begin{equation}
  Q^2 = 0, \qquad [{\cal H}, Q] = 0, \qquad \left\{Q, \Qp\right\} = {\cal H},
  \label{eq:SSQM}
\end{equation}
together with their Hermitian conjugates~\cite{witten}. In such a context, shape
invariance~\cite{genden} provides an integrability condition, yielding all the
bound state energy eigenvalues and eigenfunctions, as well as the scattering
matrix.\par
%
%
Recently, Sukhatme, Rasinariu, and Khare~\cite{sukhatme} introduced cyclic shape
invariant potentials of period $p$ in SSQM. They are characterized by the
fact that
the supersymmetric partner Hamiltonians correspond to a series of shape
invariant
potentials, which repeats after a cycle of $p$ iterations. In other words,
one may
define $p$ sets of operators $\left\{{\cal H}_{\mu}, \Qp_{\mu},
Q_{\mu}\right\}$,
$\mu=0$, 1, \ldots,~$p-1$, each satisfying the sqm(2) defining
relations~(\ref{eq:SSQM}). The operators may be written as
\begin{equation}
  {\cal H}_{\mu} = \left(\begin{array}{cc}
                 {\cal H}^{(\mu)} - {\cal E}^{(\mu)}_0 I & 0 \\
                 0 & {\cal H}^{(\mu+1)} - {\cal E}^{(\mu)}_0 I
                            \end{array}\right), \quad
  \Qp_{\mu} = \left(\begin{array}{cc}
                 0 & \Ap_{\mu} \\
                 0 & 0
                     \end{array}\right), \quad
  Q_{\mu} = \left(\begin{array}{cc}
                 0 & 0 \\
                 A_{\mu} & 0
                  \end{array}\right),     \label{eq:super-op}
\end{equation}
where
\begin{eqnarray}
  {\cal H}^{(0)} & = & \Ap_0 A_0, \nonumber \\
  {\cal H}^{(\mu)} & = & A_{\mu-1} \Ap_{\mu-1} + {\cal E}^{(\mu-1)}_0 I =
\Ap_{\mu}
          A_{\mu} + {\cal E}^{(\mu)}_0 I, \qquad \mu = 1, 2, \ldots, p,
\nonumber \\
  A_{\mu} & = & \frac{d}{dx} + W(x,b_{\mu}), \qquad \Ap_{\mu} = - \frac{d}{dx} +
  W(x,b_{\mu}), \qquad \mu = 0, 1, \ldots, p,
          \label{eq:hierarchy}
\end{eqnarray}
and ${\cal E}^{(\mu)}_0$ denotes the ground state energy of~${\cal H}^{(\mu)}$
(with ${\cal E}^{(0)}_0 = 0$). Here the superpotentials $W(x,b_{\mu})$
depend upon
some parameters $b_{\mu}$, such that $b_{\mu+p} = b_{\mu}$, and they satisfy $p$
shape invariance conditions
\begin{equation}
  W^2(x,b_{\mu}) + W'(x,b_{\mu}) = W^2(x,b_{\mu+1}) - W'(x,b_{\mu+1}) +
  \omega_{\mu}, \qquad \mu = 0, 1, \ldots, p-1,  \label{eq:shape}
\end{equation}
where $\omega_{\mu}$, $\mu=0$, 1, \ldots,~$p-1$, are some real constants.\par
%
%
{}From the solution of Eq.~(\ref{eq:shape}), one may then construct the
potentials
corresponding to the supersymmetric partners ${\cal H}^{(\mu)}$, ${\cal
H}^{(\mu+1)}$ in the usual way, i.e., $V^{(\mu)} = W^2(x, b_{\mu}) - W'(x,
b_{\mu}) +
{\cal E}^{(\mu)}_0$, $V^{(\mu+1)} = W^2(x, b_{\mu}) + W'(x, b_{\mu}) +
{\cal E}^{(\mu)}_0$. For $p=2$, Gangopadhyaya and Sukhatme~\cite{gango} obtained
such potentials as superpositions of a Calogero potential and a
$\delta$-function
singularity. For $p\ge3$, however, only numerical solutions of the shape
invariance
conditions~(\ref{eq:shape}) have been obtained~\cite{sukhatme}, so that no
analytical form of $V^{(\mu)}$ is known. In spite of this, the spectrum is
easily
derived and consists of $p$ infinite sets of equally spaced energy levels,
shifted
with respect to each other by the energies $\omega_0$, $\omega_1$,
\ldots,~$\omega_{p-1}$.\par
%
%
Since for some special choices of parameters, spectra of a similar type may be
obtained with the bosonic oscillator Hamiltonian~(\ref{eq:H_0}) acting in the
bosonic Fock space representation of ${\cal A}^{(p)}(G(N))$, one may try to
establish
a relation between the class of algebras ${\cal A}^{(p)}(G(N))$ and SSQM
with cyclic
shape invariant potentials of period~$p$.\par
%
%
In Ref.~\cite{cq99a}, we proved that the operators ${\cal H}^{(\mu)}$,
$\Ap_{\mu}$,
and $A_{\mu}$ of Eqs.~(\ref{eq:super-op}) and~(\ref{eq:hierarchy}) can be
realized
in terms of the generators of $p$ algebras ${\cal A}^{(p)}(G^{(\mu)}(N))$,
$\mu=0$, 1,
\ldots,~$p-1$, belonging to the class $\left\{{\cal A}^{(p)}(G(N))\right\}$. The
parameters of such algebras are obtained by cyclic permutations from a starting set
$\{\alpha_0, \alpha_1, \ldots, \alpha_{p-1}\}$ corresponding to ${\cal
A}^{(p)}(G^{(0)}(N)) = {\cal A}^{(p)}(G(N))$. Denoting by $N$, $\ap_{\mu}$,
$a_{\mu}$
the number, creation, and annihilation operators corresponding to the $\mu$th
algebra ${\cal A}^{(p)}(G^{(\mu)}(N))$, where $\ap_0 = \ap$, and $a_0 = a$,
we may
write the fourth relation in the algebra defining
relations~(\ref{eq:alg-def2}) as
\begin{equation}
  \left[a_{\mu}, \ap_{\mu}\right] = I + \sum_{\nu=0}^{p-1} \alpha^{(\mu)}_{\nu}
  P_{\nu}, \qquad \alpha^{(\mu)}_{\nu} \equiv \alpha_{\nu+\mu}, \qquad \mu=0, 1,
  \ldots, p-1,
\end{equation}
while the remaining relations keep the same form.\par
%
%
The realization of ${\cal H}^{(\mu)}$, $\Ap_{\mu}$, $A_{\mu}$, $\mu=0$, 1,
\ldots,~$p-1$, is then given by
\begin{eqnarray}
  {\cal H}^{(\mu)} & = & F(N+\mu) = N + \mu I + \sum_{\nu=0}^{p-1}
\beta_{\nu+\mu}
           P_{\nu} = H^{(\mu)}_0 - \case{1}{2} \sum_{\nu=0}^{p-1} \left(1 +
           \alpha^{(\mu)}_{\nu}\right) P_{\nu} + {\cal E}^{(\mu)}_0 I,
\nonumber\\
  \Ap_{\mu} & = & \ap_{\mu}, \qquad A_{\mu} = a_{\mu},
\label{eq:hierarchy-realiz}
\end{eqnarray}
where $H^{(\mu)}_0 \equiv \frac{1}{2} \left\{a^{\vphantom{\dagger}}_{\mu},
\ap_{\mu}\right\}$ is the bosonic oscillator Hamiltonian associated with ${\cal
A}^{(p)}(G^{(\mu)}(N))$, ${\cal E}^{(\mu)}_0 = \sum_{\nu=0}^{\mu-1}
\omega_{\nu}$, and the level spacings are $\omega_{\mu} = 1 + \alpha_{\mu}$. For
this result to be meaningful, the conditions $\omega_{\mu} > 0$, $\mu=0$, 1,
\ldots,~$p-1$, have to be fulfilled. When combined with the
restrictions~(\ref{eq:cond-Fock}), the latter imply that the parameters of the
starting algebra ${\cal A}^{(p)}(G(N))$ must be such that $-1 < \alpha_0 <
\lambda-1$, $-1 <
\alpha_{\mu} < \lambda - \mu -1 - \sum_{\nu=0}^{\mu-1} \alpha_{\nu}$ if
$\mu=1$, 2, $\ldots$,~$\lambda-2$, and $\alpha_{\lambda-1} = -
\sum_{\nu=0}^{\lambda-2} \alpha_{\nu}$.\par
%
%
\section{\boldmath Application to parasupersymmetric quantum mechanics of order
$p$}
\setcounter{equation}{0}

The sqm(2) superalgebra~(\ref{eq:SSQM}) is most often realized in terms of
mutually commuting boson and fermion operators. Plyushchay~\cite{plyu}, however,
showed that it can alternatively be realized in terms of only boson-like
operators,
namely the generators of the Calogero-Vasiliev algebra \gdoatwo\ (see also
Ref.~\cite{beckers97}). Such an SSQM bosonization can be performed in two
different ways, by choosing either $Q = \ap P_1$ (so that ${\cal H} = H_0 -
\frac{1}{2}(K + \kappa)$) or $Q = \ap P_0$ (so that ${\cal H} = H_0 +
\frac{1}{2}(K + \kappa)$). The first choice corresponds to unbroken SSQM
(all the
excited states are twofold degenerate while the ground state is nondegenerate
and at vanishing energy), and the second choice describes broken SSQM (all the
states are twofold degenerate and at positive energy).\par
%
%
SSQM was generalized to parasupersymmetric quantum mechanics (PSSQM) of
order two by Rubakov and Spiridonov~\cite{rubakov}, and later on to PSSQM of
arbitrary order $p$ by Khare~\cite{khare93a}. In the latter case,
Eq.~(\ref{eq:SSQM})
is replaced by
\[
Q^{p+1} = 0 \qquad ({\rm with\ } Q^p \ne 0),
\]
\[
  [{\cal H}, Q] = 0,
\]
\begin{equation}
  Q^p \Qp + Q^{p-1} \Qp Q + \cdots + Q \Qp Q^{p-1} + \Qp Q^p = 2p Q^{p-1}
{\cal H},
  \label{eq:PSSQM}
\end{equation}
and is retrieved in the case where $p=1$. The parasupercharges $Q$, $\Qp$,
and the
parasupersymmetric Hamiltonian $\cal H$ are usually realized in terms of
mutually
commuting boson and parafermion operators.\par
%
%
A property of PSSQM of order $p$ is that the spectrum of $\cal H$ is
($p+1$)-fold
degenerate above the ($p-1$)th energy level. This fact and Plyushchay's results for
$p=1$ hint at a possibility of representing $\cal H$ as a linear
combination of the
bosonic oscillator Hamiltonian $H_0$ associated with ${\cal
A}^{(p+1)}(G(N))$ and
some projection operators, as in Eq.~(\ref{eq:hierarchy-realiz}).\par
%
%
In Ref.~\cite{cq99b} (see also Refs.~\cite{cq98a,cq98b}), we proved that
PSSQM of
order
$p$ can indeed be bosonized in terms of the generators of ${\cal
A}^{(p+1)}(G(N))$
for any allowed (i.e., satisfying Eq.~(\ref{eq:cond-Fock})) values of the
algebra
parameters $\alpha_0$, $\alpha_1$, \ldots,~$\alpha_{p-1}$. For such a
purpose, we
started from ans\" atze of the type
\begin{equation}
  Q = \sum_{\nu=0}^p \sigma_{\nu} \ap P_{\nu}, \qquad {\cal H} = H_0 +
\case{1}{2}
  \sum_{\nu=0}^p r_{\nu} P_{\nu},
\end{equation}
where $\sigma_{\nu}$ and $r_{\nu}$ are some complex and real constants,
respectively, to be determined in such a way that Eq.~(\ref{eq:PSSQM}) is
fulfilled.
We found that there are $p+1$ families of solutions, which may be
distinguished by
an index $\mu \in
\{0, 1, \ldots, p\}$ and from which we may choose the following representative
solutions
\begin{eqnarray}
  Q_{\mu} & = & \sqrt{2} \sum_{\nu=1}^p \ap P_{\mu+\nu}, \nonumber\\
  {\cal H}_{\mu} & = &  N + \case{1}{2} (2\gamma_{\mu+2} + r_{\mu+2} - 2p +
3) I +
          \sum_{\nu=1}^p (p + 1 - \nu) P_{\mu+\nu},  \label{eq:PSSQM-sol}
\end{eqnarray}
where
\begin{equation}
  r_{\mu+2} = \frac{1}{p} \left[(p-2) \alpha_{\mu+2} + 2 \sum_{\nu=3}^p
(p-\nu+1)
  \alpha_{\mu+\nu} + p (p-2)\right].
\end{equation}
\par
%
%
The eigenvectors of ${\cal H}_{\mu}$ are the states~(\ref{eq:vectors}) and the
corresponding eigenvalues are easily found. All the energy levels are equally
spaced. For $\mu=0$, PSSQM is unbroken, otherwise it is broken with a
($\mu+1$)-fold degenerate ground state. All the excited states are ($p+1$)-fold
degenerate. For $\mu=0$, 1, \ldots,~$p-2$, the ground state energy may be
positive,
null, or negative depending on the parameters, whereas for $\mu = p-1$ or
$p$, it is
always positive.\par
%
%
Khare~\cite{khare93a} showed that in PSSQM of order $p$, $\cal H$ has in
fact $2p$
(and not only two) conserved parasupercharges, as well as $p$ bosonic
constants. In
other words, there exist $p$ independent operators $Q_r$, $r=1$, 2, \ldots,~$p$,
satisfying with $\cal H$ the set of equations~(\ref{eq:PSSQM}), and $p$ other
independent operators $I_t$, $t=2$, 3, \ldots,~$p+1$, commuting with $\cal
H$, as
well as among themselves. In Ref.~\cite{cq99b}, we obtained a realization of all
such operators in terms of the ${\cal A}^{(p+1)}(G(N))$ generators.\par
%
%
As a final point, let us note that there exists an alternative approach to
PSSQM of
order $p$, which was proposed by Beckers and Debergh~\cite{beckers90}, and
wherein the multilinear relation in Eq.~(\ref{eq:PSSQM}) is replaced by the
cubic
equation
\begin{equation}
  \left[Q, \left[\Qp, Q\right] \right] = 2Q {\cal H}.  \label{eq:cubic}
\end{equation}
In Ref.~\cite{cq98a}, we proved that for $p=2$, this PSSQM algebra can only be
realized by those ${\cal A}^{(3)}(G(N))$ algebras that simultaneously bosonize
Rubakov-Spiridonov-Khare PSSQM algebra.\par
%
%
\section{Application to pseudosupersymmetric quantum mechanics}
\setcounter{equation}{0}

Pseudosupersymmetric quantum mechanics (pseudoSSQM) was introduced by
Beckers, Debergh, and Nikitin~\cite{beckers95} in a study of relativistic vector
mesons interacting with an external constant magnetic field. In the
nonrelativistic
limit, their theory leads to a pseudosupersymmetric oscillator Hamiltonian,
which
can be realized in terms of mutually commuting boson and pseudofermion
operators,
where the latter are intermediate between standard fermion and $p=2$
parafermion operators.\par
%
%
It is then possible to formulate a pseudoSSQM~\cite{beckers95},
characterized by a
pseudosupersymmetric Hamiltonian $\cal H$ and pseudosupercharge operators $Q$,
$Q^{\dagger}$, satisfying the relations
\begin{equation}
  Q^2 = 0, \qquad [{\cal H}, Q] = 0, \qquad Q Q^{\dagger} Q = 4 c^2 Q {\cal H},
  \label{eq:pseudoSSQM}
\end{equation}
and their Hermitian conjugates, where $c$ is some real constant. The first two
relations in Eq.~(\ref{eq:pseudoSSQM}) are the same as those occurring in SSQM,
whereas the third one is similar to the multilinear relation valid in PSSQM
of order
two. Actually, for $c=1$ or 1/2, it is compatible with Eq.~(\ref{eq:PSSQM}) or
(\ref{eq:cubic}), respectively.\par
%
%
In Ref.~\cite{cq99b}, we proved that pseudoSSQM can be bosonized in two
different
ways in terms of the generators of ${\cal A}^{(3)}(G(N))$ for any allowed
values of
the parameters $\alpha_0$, $\alpha_1$. This time, we started from the ans\" atze
\begin{equation}
  Q = \sum_{\nu=0}^2 \left(\xi_{\nu} a + \eta_{\nu} a^{\dagger}\right) P_{\nu},
  \qquad {\cal H} = H_0 + \case{1}{2} \sum_{\nu=0}^2 r_{\nu} P_{\nu},
\end{equation}
and determined the complex constants $\xi_{\nu}$, $\eta_{\nu}$, and the
real ones
$r_{\nu}$ in such a way that Eq.~(\ref{eq:pseudoSSQM}) is fulfilled.\par
%
%
The first type of bosonization corresponds to three families of two-parameter
solutions, labelled by an index $\mu \in \{0, 1, 2\}$,
\begin{eqnarray}
  Q_{\mu}(\eta_{\mu+2}, \varphi) & = & \left(\eta_{\mu+2} a^{\dagger} +
e^{{\rm i}
          \varphi}\sqrt{4 c^2 - \eta_{\mu+2}^2}\, a\right) P_{\mu+2},
\nonumber \\
  {\cal H}_{\mu}(\eta_{\mu+2}) & = & N + \case{1}{2} (2 \gamma_{\mu+2} +
          r_{\mu+2} - 1) I + 2 P_{\mu+1} + P_{\mu+2},  \label{eq:pseudoSSQM-sol}
\end{eqnarray}
where $0 < \eta_{\mu+2} < 2 |c|$, $0 \le \varphi < 2\pi$, and
\begin{equation}
  r_{\mu+2} = \frac{1}{2c^2} (1 + \alpha_{\mu+2}) \left(|\eta_{\mu+2}|^2 - 2
  c^2\right).
\end{equation}
Choosing for instance $\eta_{\mu+2} = \sqrt{2} |c|$, and $\varphi = 0$, hence
$r_{\mu+2} = 0$ (producing an overall shift of the spectrum), we obtain
\begin{eqnarray}
  Q_{\mu} & = & c \sqrt{2} \left(a^{\dagger} + a\right) P_{\mu+2}, \nonumber \\
  {\cal H}_{\mu} & = & N + \case{1}{2} (2 \gamma_{\mu+2} - 1) I + 2 P_{\mu+1} +
           P_{\mu+2}.  \label{eq:pseudoSSQM-solbis}
\end{eqnarray}
A comparison between Eq.~(\ref{eq:pseudoSSQM-sol}) or
(\ref{eq:pseudoSSQM-solbis}) and Eq.~(\ref{eq:PSSQM-sol}) shows that the
pseudosupersymmetric and $p=2$ parasupersymmetric Hamiltonians coincide, but
that the corresponding charges are of course different. The conclusions
relative to
the spectrum and the ground state energy are therefore the same as in Sec.~4.
\par
%
%
The second type of bosonization corresponds to three families of one-parameter
solutions, again labelled by an index $\mu \in \{0, 1, 2\}$,
\begin{eqnarray}
  Q_{\mu} & = & 2 |c| a P_{\mu+2}, \nonumber \\
  {\cal H}_{\mu}(r_{\mu}) & = & N + \case{1}{2} (2 \gamma_{\mu+2}
          - \alpha_{\mu+2}) I + \case{1}{2} (1 - \alpha_{\mu+1} +
\alpha_{\mu+2}
          + r_{\mu}) P_{\mu} + P_{\mu+1},
\end{eqnarray}
where $r_{\mu} \in {\rm R}$ changes the Hamiltonian spectrum in a significant
way. We indeed find that the levels are equally spaced if and only if $r_{\mu} =
(\alpha_{\mu+1} - \alpha_{\mu+2} + 3)\, {\rm mod}\, 6$. If $r_{\mu}$ is small
enough, the ground state is nondegenerate, and its energy is negative for
$\mu=1$,
or may have any sign for $\mu = 0$ or~2. On the contrary, if $r_{\mu}$ is large
enough, the ground state remains nondegenerate with a vanishing energy in the
former case, while it becomes twofold degenerate with a positive energy in the
latter. For some intermediate $r_{\mu}$ value, one gets a two or threefold
degenerate ground state with a vanishing or positive energy, respectively.\par
%
%
\section{Application to orthosupersymmetric quantum mechanics of order two}
\setcounter{equation}{0}

Mishra and Rajasekaran~\cite{mishra} introduced order-$p$ orthofermion operators
by replacing the Pauli exclusion principle by a more stringent one: an orbital
state shall not contain more than one particle, whatever be the spin direction.
The wave function is thus antisymmetric in spatial indices alone with the
order of the spin indices frozen.\par
%
%
Khare, Mishra, and Rajasekaran~\cite{khare93b} then developed
orthosupersymmetric quantum mechanics (OSSQM) of arbitrary order $p$ by
combining boson operators with orthofermion ones, for which the spatial indices
are ignored. OSSQM is formulated in terms of an orthosupersymmetric Hamiltonian
$\cal H$, and $2p$ orthosupercharge operators $Q_r$, $Q_r^{\dagger}$, $r =
1$, 2,
\ldots,~$p$, satisfying the relations
\begin{equation}
  Q_r Q_s = 0, \qquad [{\cal H}, Q_r] = 0, \qquad Q_r Q_s^{\dagger} +
\delta_{r,s}
  \sum_{t=1}^p Q_t^{\dagger} Q_t = 2 \delta_{r,s} {\cal H},  \label{eq:OSSQM}
\end{equation}
and their Hermitian conjugates, where $r$ and $s$ run over 1, 2,
\ldots,~$p$.\par
%
%
In Ref.~\cite{cq99b}, we proved that OSSQM of order two can be bosonized in
terms
of the generators of some well-chosen ${\cal A}^{(3)}(G(N))$ algebras. As ans\"
atze, we used the expressions
\begin{equation}
  Q_1 = \sum_{\nu=0}^2 \left(\xi_{\nu} a + \eta_{\nu} a^{\dagger}\right)
P_{\nu},
  \qquad Q_2 = \sum_{\nu=0}^2 \left(\zeta_{\nu} a + \rho_{\nu}
a^{\dagger}\right)
  P_{\nu}, \qquad {\cal H} = H_0 + \case{1}{2} \sum_{\nu=0}^2 r_{\nu}
P_{\nu},
\end{equation}
and determined the complex constants $\xi_{\nu}$, $\eta_{\nu}$, $\zeta_{\nu}$,
$\rho_{\nu}$, and the real ones $r_{\nu}$ in such a way that
Eq.~(\ref{eq:OSSQM}) is
fulfilled. We found two families of two-parameter solutions, labelled by
$\mu \in
\{0 ,1\}$,
\begin{eqnarray}
  Q_{1,\mu}(\xi_{\mu+2}, \varphi) & = & \xi_{\mu+2} a P_{\mu+2} + e^{{\rm i}
         \varphi} \sqrt{2 - \xi_{\mu+2}^2}\, a^{\dagger} P_{\mu}, \nonumber \\
  Q_{2,\mu}(\xi_{\mu+2}, \varphi) & = & - e^{-{\rm i} \varphi} \sqrt{2 -
         \xi_{\mu+2}^2}\, a P_{\mu+2} + \xi_{\mu+2} a^{\dagger} P_{\mu},
         \nonumber\\
  {\cal H}_{\mu} & = & N + \case{1}{2} (2 \gamma_{\mu+1} - 1) I + 2 P_{\mu} +
         P_{\mu+1},  \label{eq:OSSQM-sol}
\end{eqnarray}
where $0 < \xi_{\mu+2} \le \sqrt{2}$ and $0 \le \varphi <2\pi$, provided the
algebra parameter $\alpha_{\mu+1}$ is taken as $\alpha_{\mu+1} = -1$. As a
matter
of fact, the absence of a third family of solutions corresponding to
$\mu=2$ comes
from the incompatibility of this condition (i.e., $\alpha_0 = -1$) with
conditions~(\ref{eq:cond-Fock}).\par
%
%
The orthosupersymmetric Hamiltonian $\cal H$ in Eq.~(\ref{eq:OSSQM-sol}) is
independent of the parameters $\xi_{\mu+2}$, $\varphi$. All the levels of its
spectrum are equally spaced. For $\mu=0$, OSSQM is broken: the levels are
threefold
degenerate, and the ground state energy is positive. On the contrary, for
$\mu=1$,
OSSQM is unbroken: only the excited states are threefold degenerate, while the
nondegenerate ground state has a vanishing energy. Such results agree with the
general conclusions of Ref.\ \cite{khare93b}.\par
%
%
{}For $p$ values greater than two, the OSSQM algebra~(\ref{eq:OSSQM}) becomes
rather complicated because the number of equations to be fulfilled increases
considerably. A glance at the 18 independent conditions for $p=3$ led us to the
conclusion that the ${\cal A}^{(4)}(G(N))$ algebra is not rich enough to contain
operators satisfying Eq.~(\ref{eq:OSSQM}). Contrary to what happens for PSSQM,
for OSSQM the $p=2$ case is therefore not representative of the general one.\par
%
%
\section{Conclusion}

In this communication, we showed that the $S_2$-extended oscillator algebra,
which was introduced in connection with the two-particle Calogero model, can be
extended to the whole class of $C_{\lambda}$-extended oscillator algebras \alg,
where $\lambda \in \{2,3, \ldots\}$, and $\alpha_0$, $\alpha_1$,
\ldots,~$\alpha_{\lambda-2}$ are some real parameters. In the same way, the GDOA
realization of the former, known as the Calogero-Vasiliev algebra, is
generalized to
a class of GDOAs \gdoa, where $\lambda \in \{2,3, \ldots\}$, for which one can
define a bosonic oscillator Hamiltonian $H_0$, acting in the bosonic Fock space
representation.\par
%
%
{}For $\lambda \ge 3$, the spectrum of $H_0$ has a very rich structure in
terms of
the algebra parameters $\alpha_0$, $\alpha_1$, \ldots,~$\alpha_{\lambda-2}$.
This can be exploited to provide an algebraic realization of SSQM with
cyclic shape
invariant potentials of period~$\lambda$, a bosonization of PSSQM of order $p =
\lambda-1$, and, for $\lambda=3$, a bosonization of pseudoSSQM and OSSQM of
order two.\par
%
%

\end{document}